\documentclass[article,twocolumn,floatfix]{revtex4}

\usepackage{graphicx} 
\usepackage{epstopdf}
\usepackage{subfig}
\usepackage{array}
\usepackage{color}

\newcommand{\hf}{\frac{1}{2}}
\newcommand{\be}{\begin{equation}}
\newcommand{\ee}{\end{equation}}
\newcommand{\bea}{\begin{eqnarray}}
\newcommand{\eea}{\end{eqnarray}}

\begin{document}
\title{Missing data outside the detector range: application to continuous variable entanglement verification and quantum cryptography}
\author{Megan R.~Ray and S.J. van Enk}
\affiliation{Department of Physics and Oregon Center for Optics,\\
University of Oregon, Eugene, OR 97403}
\begin{abstract}
In continuous-variable quantum information processing detectors are necessarily coarse grained and of finite range. We discuss how especially the latter feature is a bug and 
may easily lead to overoptimistic estimates of entanglement and of security, when missed data outside the detector range are ignored. We show that entropic separability or security criteria are much superior to variance-based criteria for mitigating the negative effects of this bug.
 \end{abstract}
\maketitle

Secure quantum key distribution requires the ability of the sender and receiver to use their measurement results to prove the presence of entanglement in the (effective) quantum state distributed between them \cite{curty}. The problem of how to decide whether a given state is separable or entangled continues to be an area of active research \cite{guhne2009entanglement}. In almost all applications it is important to avoid concluding there is entanglement when there is, in fact, none.

There has been great interest in continuous-variable (CV) entanglement for quantum information processing (for example, for a handful of very recent experiments, see Refs.~\cite{zhong2012efficient,jouguet2012experimental,yukawa2012generating,threephoton,madsen2012continuous}). One reason is that in optics CV states, such as coherent states or two-mode squeezed states, are easier to generate than single-photon polarization states. Moreover, by using continuous degrees of freedom (e.g., quadratures, or frequency) rather than polarization, a single photon can carry more information than just one (qu)bit.

In every entanglement verification or quantum cryptography experiment one has to deal with missing data due to imperfect detectors. For missed counts {\em inside} the detector range we typically assume that the missing data would follow the same statistics as the recorded data. 
(When 
Bell inequalities are used to eliminate hidden-variable theories no such assumption may be used, thus leading to the so-called detection loophole; when using the same inequalities for entanglement verification the assumption is typically warranted \cite{vanenk}.) Here, in contrast, we worry about missed counts from {\em outside} one's detector range. The assumption that those counts would follow the same statistics as the recorded data is meaningless. We will show that it depends on what criterion one uses to verify entanglement (or to prove cryptographic security) how one should take into account missing data of this type.

The best known and easiest to calculate separability criterion developed for detecting discrete variable bipartite entanglement is the positive partial transpose (PPT) criterion \cite{Peres}.  The PPT criterion has been extended to bipartite CV systems by Shchukin and Vogel \cite{vogel} and by Miranowicz {\em et al.} \cite{miranowicz}, which encompasses previous CV criteria by Duan {\em et al.} \cite{Duan}, Simon\cite{Simon}, Raymer {\em et al.}\cite{raymer}, Mancini {\em et al.}\cite{mancini}, and others. For example the Mancini-Giovanetti-Vitali-Tombesi (MGVT) criterion states that if $X_-=x_a-x_b$ and $P_+=p_a+p_b$, where $x$ and $p$ are dimensionless scaled position and momentum variables for particles $a$ and $b$, with $[x,p]=i$, then for all separable states
\begin{equation}\label{cond1}
\sigma^2[X_-]\sigma^2[P_+]\ge1.
\end{equation}
Violation of this condition on variances means that
the underlying state is verifiably entangled. 

To use this criterion to verify continuous variable entanglement of a state using experimental data, one must take into account how the detectors are used to measure the state, and possibly modify the criterion accordingly.  The first thing that must be considered is the coarse grained nature of the measurements.  While the variable being measured is continuous, our detectors have finite resolution and our data is binned. Instead of measuring $\hat{x}$ and $\hat{p}$ we are measuring the observables
\begin{eqnarray}
\hat{x}^\Delta&=&\sum_{k=1}^D \int_{(k-1/2)\Delta}^{(k+1/2)\Delta}dx \,x_k  |x\rangle \langle x|\\
\hat{p}^\delta&=&\sum_{l=1}^D  \int_{(l-1/2)\delta}^{(l+1/2)\delta}dp \,p_l|p\rangle \langle p|
\end{eqnarray}
where $x_k=k\Delta+x_0$ and $p_l=l\delta+p_0$, and
where $\Delta$ and $\delta$ are the resolutions of the $x$ and $p$ detectors, respectively. We also assumed the number of bins, $D$, to be the same for $x$ and $p$ measurements.  

Tasca {\em et al.} \cite{tasca} modified the MGVT criterion for such coarse grained measurements.  The variances we need to calculate are now given by
\begin{eqnarray}
\sigma^2[X_{-}^\Delta]&=&\langle(\hat{x}_{a}^\Delta-\hat{x}_{b}^\Delta)^2\rangle-\langle(\hat{x}_{a}^\Delta-\hat{x}_{b}^\Delta)\rangle^2\nonumber\\
\sigma^2[P_+^\delta]&=&\langle(\hat{p}_{a}^\delta+\hat{p}_{b}^\delta)^2\rangle-\langle(\hat{p}_{a}^\delta+\hat{p}_{b}^\delta)\rangle^2,
\end{eqnarray}
and all separable states satisfy \cite{tasca}
\begin{equation}\label{cond2}
\left(\sigma^2[X^\Delta_-]+\frac{\Delta^2}{12}\right)\left(\sigma^2[P^\delta_+]+\frac{\delta^2}{12}\right)\ge1,
\end{equation}
and so a violation proves entanglement. This condition is harder to violate than (\ref{cond1}). That is, the binning of the data 
loses information, and makes it harder to verify entanglement.

The correction for coarse graining alone is not sufficient to properly verify entanglement experimentally. An additional correction must be made to take into account the finite detection range of the detectors.  In general the wavefunction is not zero outside the detection range, so there is some probability of ``missed counts''-- events when the particle arrives at the detector but is not detected because it falls outside the detection range. (This is different than not being detected because of transmission loss or detector inefficiency.) 

To illustrate the importance of being earnest about the missed counts, let us first consider a pure {\em separable}  gaussian bipartite state, shared between Alice and Bob, of the form
\begin{equation}
\Psi(p_a,p_b)\propto \exp\left(\frac{-p_a^2}{s}\right)\exp\left(\frac{-p_b^2}{s}\right),
\end{equation}
with $s$ chosen such that
the particles' wavefunction is well localized in the variables $p_a$ and $p_b$. Hence, the value of $P^\delta_+$ is quite sharply defined, but, as a consequence, the probability distribution for the complementary variables $x_a$ and $x_b$  is broad and $X^{\Delta}_-$ is not sharply defined.
For simplicity we will assume in our examples that Alice and Bob are using detectors of the same resolution and range for both $x_{a,b}$ and $p_{a,b}$.
Since the probability distribution in $x_{a,b}$ extends well beyond their detection range, the probability of Alice and Bob both detecting a particle when measuring in the $x_{a,b}$ basis will be small. For, e.g., $s=3$ and a detection range of $[-2,2]$ this probability is only about $8.6\%$, whereas in the $p_{a,b}$ basis it is essentially $100\%$. From now on, we always fix Alice's and Bob's detector ranges to $[-2,2]$.

Now consider a mixed state consisting of a 50/50 mixture of the separable state we just examined and a similar separable state equally sharply peaked in $x_a$ and $x_b$.  This mixed state will have sharp  features in both $x$ and $p$, while most of the broad background distribution falls outside the detection window. The probability of detecting both particles is only $54.3\%$ for either basis. Ignoring the missed counts, $X^{\Delta}_-$ and $P^{\delta}_+$ are mostly sharply defined, and look similar to what the results would be for an entangled EPR-like state.  After measuring the state with 32 bin resolution detectors, naively normalizing the detected data to 100\% as in Figure \ref{wrongway}, and calculating $(\sigma^2[P^\delta_+]+\frac{\delta^2}{12})(\sigma^2[X^\Delta_-]+\frac{\Delta^2}{12})\approx .05<1$, the coarse grained MGVT criterion will ``verify entanglement" even though the underlying state clearly is separable.

\begin{figure*}
\begin{center}
\begin{tabular}{cccc}
\subfloat{\includegraphics[width=6cm]{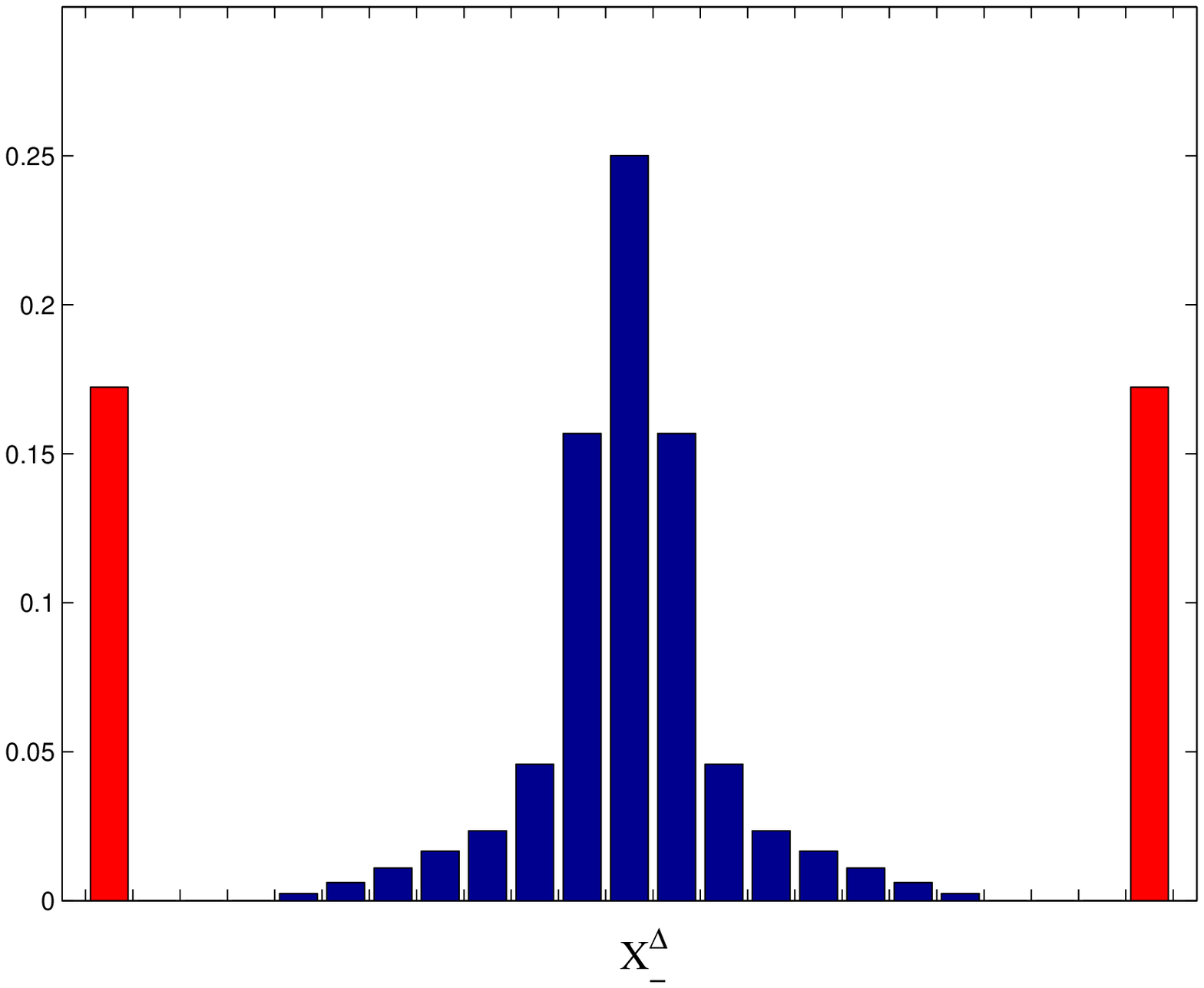}\label{cartoon}} & 
\subfloat{\includegraphics[width=6cm]{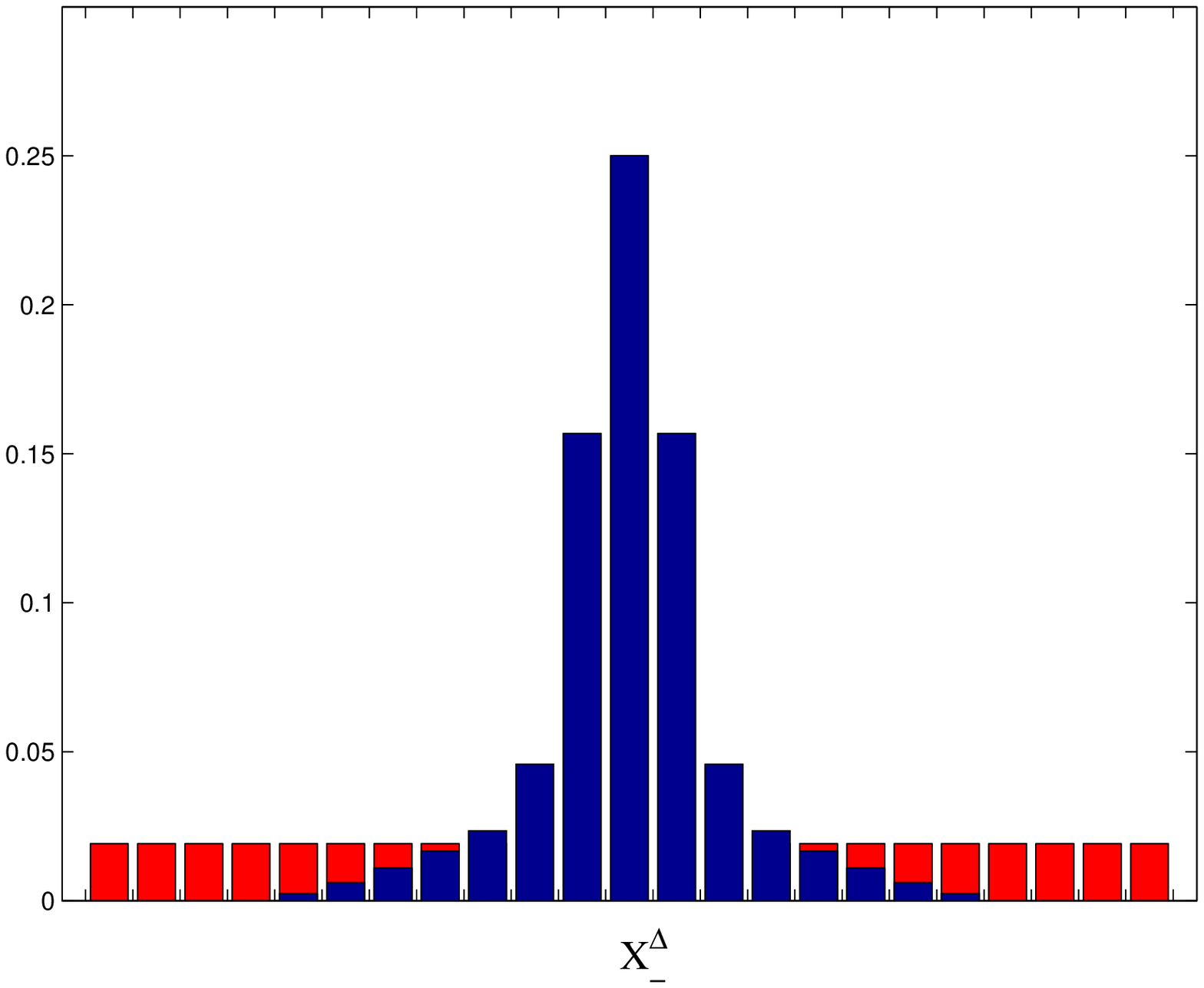}\label{cartoone}}\\
\end{tabular}
\caption{Left: Cartoon showing how to deal with missing data (red) outside our detector range for a {\em variance}-based entanglement or security criterion. After having determined a cutoff range beyond which we expect no counts, missed data are assigned to the outside bins within the cutoff range. Right: Same for an {\em entropic} entanglement or security criterion: missed data are as uniformly distributed as possible within the cutoff range.}
\end{center}
\end{figure*}

And so we must account for the missed data in some way.  To do this we will assume that we know what percentage of counts are missed (this information could be obtained by using a detector of lower resolution but much broader range) and we will also assume that we can set a cutoff beyond which no counts would occur even if one were to detect in that region. We will add additional bins to $X^{\Delta}_-$ and $P^{\delta}_+$ that correspond to values that could have been measured outside our detector range, but inside the our cutoff.   In order to properly verify entanglement we must then add the missed counts to our data in the worst way possible so that we will not be led to believe that we have entanglement (or security) when we do not.  For the MGVT criterion this means we should maximize the variance of  $X^{\Delta}_-$ and $P^{\delta}_+$, and so we add the missed counts to the outermost bins of $X^{\Delta}_-$ and $P^{\delta}_+$ (the weighting depends on the experimental data and the cutoffs. For symmetric experimental data and symmetric cutoffs, half of the missed counts goes into each of the outermost bins, as depicted in Figure \ref{cartoon}.)

\begin{figure}	  
\includegraphics[width=6cm]{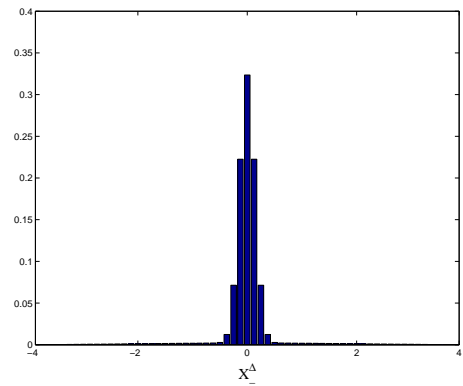}\label{wrongway}
\caption{$X^{\Delta}_-$ for mixed separable state, where the missed counts outside the detector range have  been ignored (the graph for $P^{\delta}_+$ would look identical). 
We would falsely conclude we have entanglement. Note the low barely visible background level: in order to properly reach conclusions about entanglement, we would need to know how far those background counts extend outside our detector range.}\label{wrongway}
\end{figure}

To attempt to verify entanglement for the mixed separable state example, we find that by setting cutoffs at, say, -50 and 50 for the detectors of both Bob and Alice, they would fail to both detect particles less than $10^{-18}$ percent of the time.  The choice of cutoff corresponds to adding 1536  bins for $X^{\Delta}_-$ and $P^{\delta}_+$, in the outermost of which we place the missing data. The variances increase so much we no longer come close to concluding that we have entanglement.

\begin{figure*}
\begin{center}
\begin{tabular}{cccc}
\subfloat{\includegraphics[width=6cm]{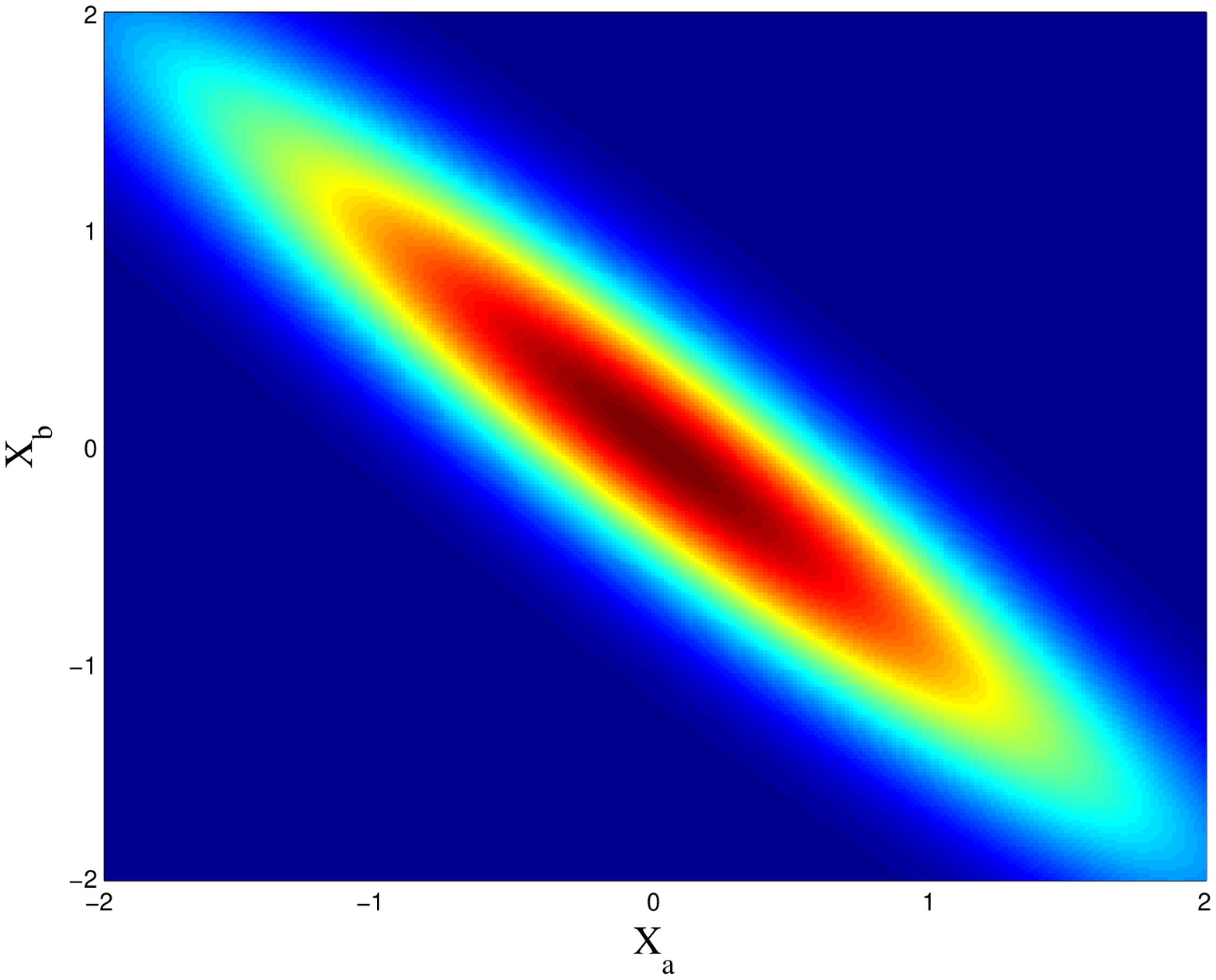}\label{epr}} & 
\subfloat{\includegraphics[width=6cm]{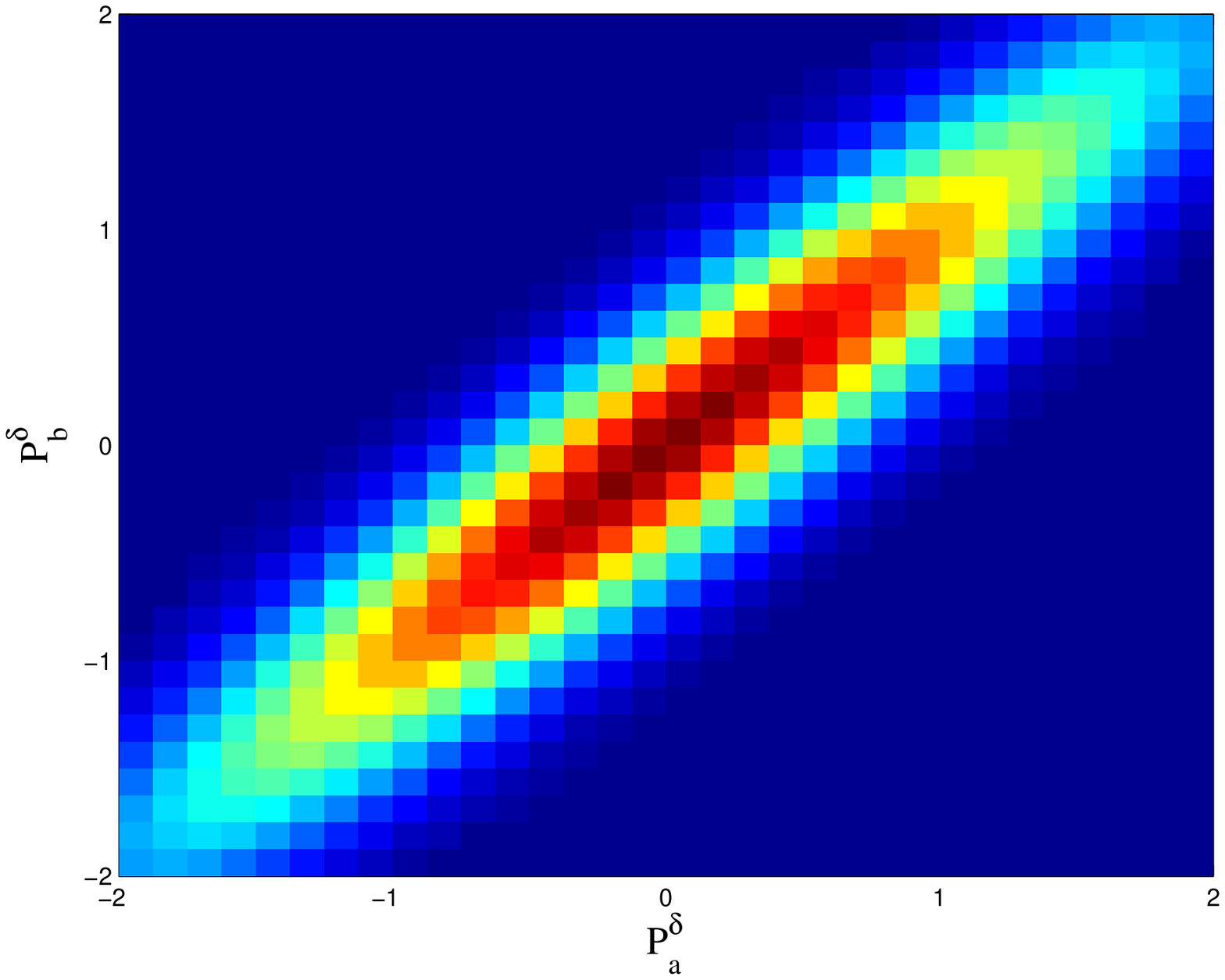}}\\
\end{tabular}
\caption{Left: $|\Psi(x_a,x_b)|^2$ for the smoothed EPR state (\ref{smEPR}) with $\bar{n}=1$; Right: $|\Psi(p_a,p_b)|^2$ for the same smoothed EPR state as measured by detectors with a 32 bin resolution}
\end{center}
\end{figure*}

Because of the necessity of adding the missing data in the worst way and the strong dependence of the variance on the cutoff and amount of missed counts, the MGVT criterion will often fail to verify entanglement when it is, in fact, present.  For example, consider an {\em entangled} gaussian state (``smoothed EPR" \cite{englert})
\bea\label{smEPR}
\Psi(x_a,x_b)\propto\exp\left(2\sqrt{\bar{n}(\bar{n}+1)}x_a x_b-(\bar{n}+\hf)(x_a^2+x_b^2)\right)
\eea
where for $\bar{n}=0$ the state is separable and in the limit ${\bar{n} \to \infty}$ we have the original EPR state. For $\bar{n}=1$ and a detection range of [-2,2] (see Figure \ref{epr}) the probability of both parties detecting a particle is about $86.9\%$ For this case, without even adding extra bins to $X^{\Delta}_-$ and $P^{\delta}_+$ and simply putting the missing data into the outermost of the existing bins inside the detector range, we already fail to verify the entanglement present in the state with the MGVT criterion, no matter what the cutoff would be. 
Other variance-based criteria fare equally badly.
 
A better choice is to use a criterion that does not depend as strongly on the location of the cutoff or the probability of missed counts. Instead of using a variance-based criterion we will now look at an entropic criterion.  Continuous variable separability criteria have been developed using Shannon, Tsallis, and Renyi entropies \cite{walborn,Saboia2011}.  We focus on the latter. The Renyi entropy of order $\alpha$ for a binned probability distribution is defined as 
\begin{equation}
H_\alpha[B^{\delta b}]=\frac{1}{1-\alpha}\ln(\sum_k (B^{\delta b}_k)^\alpha)
\end{equation}
It has been shown in Ref.~\cite{Saboia2011} that all separable states satisfy
\begin{equation}
H_{\alpha}[X_{-}^\Delta] + H_{\beta}[P_{+}^\delta] +\frac{1}{2}\left(\frac{\ln\alpha}{1-\alpha} +\frac{\ln\beta}{1-\beta}\right)-\ln\frac{2\pi}{\Delta \delta}\geq 0
\end{equation}
for $1/\alpha +1/\beta=2$. 
So if
this equality is violated
for {\em any} such constrained pair of values $\alpha$ and $\beta$   the underlying state is verifiably entangled. The criterion is optimized by minimizing the left-hand side over the allowed values for $\alpha,\beta$.

To deal with the missed counts when using this criterion we again add additional bins to $X^{\Delta}_-$ and $P^{\delta}_+$ that correspond to values that could have been measured outside our detector range, but inside a cutoff. We then must add the missing data such that it maximizes the Renyi entropy. Since the Renyi entropy is maximized by a uniform distribution, we will add the missed counts to the new empty bins and the existing bins with few counts to make the distribution as uniform as possible, as shown in cartoon form in Figure \ref{cartoone}. We then optimize our criterion and hope we verify entanglement.

For a smoothed EPR state with $\bar{n}=1$, Alice and Bob would only fail to both detect a particle less than $10^{-12}$ percent of the time in the range $[-10,10]$, so they could choose that as their cutoff range. They add 256 bins to $X^{\Delta}_-$ and $P^{\delta}_+$ and distribute the missing data in the most uniform way possible.  After doing this and optimizing the criterion, they do indeed verify entanglement. (Recall that the variance criterion always fails for this case.)

\begin{figure}[ht]
\includegraphics[width=6cm]{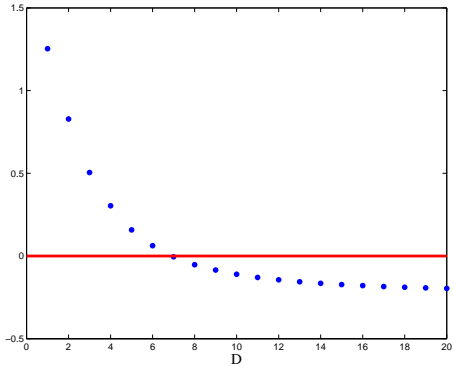}\caption{Optimized Renyi criterion for the smooted EPR state with $\bar{n}=1$ as a function of the number of bins  inside the detection range. One needs a minimum number of 7 bins to verify entanglement.}\label{nbin2}\label{reso}
 \end{figure}

The Renyi entropy criterion depends on the size of the bins both in the explicit bin width term and the Renyi entropy terms. Fixing the detection range and cutoff, and varying the number of bins $D$ of our detectors, we see in Figure \ref{reso} that in this instance entanglement will be verified if we include at least 7 bins. If we have  fewer bins, we throw too much information away and can not be sure we have entanglement.

\begin{figure}[ht]
\includegraphics[width=6cm]{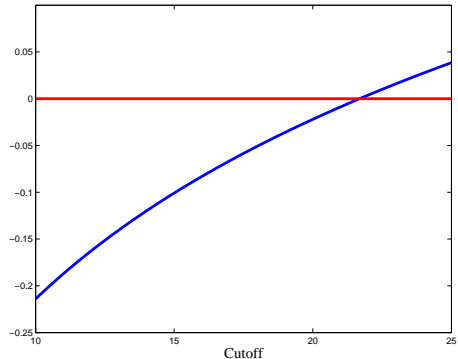}
\caption{Optimized criterion for the smooted EPR state with $\bar{n}=1$  as a function of the ``no particle cutoff'' $x$, which assumes  no detection events would occur outside the interval $[-x, x]$). One needs $x<21.5$ in order to be able to verify entanglement.}\label{nocountcutoff} \end{figure}

The strength of the Renyi entropy criterion is its smaller sensitivity to the location of the cutoff. In Figure \ref{nocountcutoff} we see that entanglement will be verified for the $\bar{n}=1$ smoothed EPR state as long as the cutoff range does not exceed $[-21.5,21.5]$, where we recall once again that the variance criterion would always fail, no matter what cutoff.

In conclusion, then, we discussed a difficulty that arises  in entanglement verification (or quantum cryptography) experiments for continuous variable systems, which does not seem to have been discussed in the literature yet: missing data {\em outside} one's detector range. We showed how to take those missing data into account, by distributing them over the outside range in the worst possible way, given the criterion one uses to verify entanglement (or prove security), as schematically pictured in Figs.~\ref{cartoon} and \ref{cartoone}. As a consequence,  entropic entanglement (security) criteria turn out to be much more forgiving than are variance-based criteria.

\bibliography{missingdataL}

\end{document}